\documentclass[twocolumn,graphicx,superscriptaddress,showkeys,floatfix]{revtex4-1} %aip

\usepackage{amsfonts}
\usepackage{graphicx}
\usepackage{bm}
\usepackage{dcolumn}
\usepackage{amssymb}
\usepackage[colorlinks,linkcolor=blue,citecolor=blue]{hyperref}
\usepackage{epsfig}
\usepackage{multirow}
\usepackage{amsmath}
\usepackage{hhline}
\usepackage{mathptmx}
\usepackage{eucal}
\usepackage{color}
\usepackage{epstopdf}
\usepackage{natbib}
\usepackage{textcomp}

\begin{document}
\title{Magnetically-driven colossal supercurrent enhancement in InAs nanowire Josephson junctions}
%\keywords{Nanowire, InAs, Josephson effect, Supercurrent enhancement, Topological transition}

\author{J. Paajaste}
%\email{NEW email of Jonna}
\affiliation{NEST, Istituto Nanoscienze-CNR and Scuola Normale Superiore, I-56127 Pisa, Italy}

\author{E. Strambini}
\email{elia.strambini@sns.it}
\affiliation{NEST, Istituto Nanoscienze-CNR and Scuola Normale Superiore, I-56127 Pisa, Italy}

\author{M. Amado}
\affiliation{NEST, Istituto Nanoscienze-CNR and Scuola Normale Superiore, I-56127 Pisa, Italy}
\affiliation{Materials Science and Metallurgy, University of Cambridge CB3 OFS, United Kingdom}

\author{S. Roddaro}
%\email{roddaro@sns.it}
\affiliation{NEST, Istituto Nanoscienze-CNR and Scuola Normale Superiore, I-56127 Pisa, Italy}

\author{P. San-Jose}
\affiliation{Instituto de Ciencia de Materiales de Madrid, Consejo Superior de Investigaciones Cient\'{\i}ficas (ICMM-CSIC), Sor Juana In\'{e}s de la Cruz 3, 28049 Madrid, Spain}

\author{R. Aguado}
\affiliation{Instituto de Ciencia de Materiales de Madrid, Consejo Superior de Investigaciones Cient\'{\i}ficas (ICMM-CSIC), Sor Juana In\'{e}s de la Cruz 3, 28049 Madrid, Spain}

\author{F. S. Bergeret}
\affiliation{Centro de Fisica de Materiales (CFM-MPC), Centro Mixto CSIC-UPV/EHU, E-20018 San Sebastian, Spain}
\affiliation{Donostia International Physics Center (DIPC), E-20018 San Sebastian, Spain}

\author{D. Ercolani}
\affiliation{NEST, Istituto Nanoscienze-CNR and Scuola Normale Superiore, I-56127 Pisa, Italy}

\author{L. Sorba}
\affiliation{NEST, Istituto Nanoscienze-CNR and Scuola Normale Superiore, I-56127 Pisa, Italy}

\author{F. Giazotto}
\email{francesco.giazotto@sns.it}
\affiliation{NEST, Istituto Nanoscienze-CNR and Scuola Normale Superiore, I-56127 Pisa, Italy}

\begin{abstract}
%The abstract is typically 150 words and is unreferenced; it contains a brief account of the background and rationale of the work, followed by a statement of the main conclusions introduced by the phrase "Here we show" or some equivalent.
\textbf{The Josephson effect is a fundamental quantum phenomenon consisting in the appearance of a dissipationless supercurrent in a weak link between two superconducting (S) electrodes. While the mechanism leading to the Josephson effect is quite general, i.e., Andreev reflections at the interface between the S electrodes and the weak link, the precise physical details and topology of the junction drastically modify the properties of the supercurrent.  Specifically, a strong enhancement of the critical supercurrent $I_C$ is expected to occur when the topology of the junction allows the emergence of Majorana bound states. Here we report charge transport measurements in mesoscopic Josephson junctions formed by InAs nanowires and Ti/Al superconducting leads. Our main observation is a colossal enhancement of the critical supercurrent induced by an external magnetic field applied perpendicular to the substrate. This striking and anomalous supercurrent enhancement cannot be ascribed to any known conventional phenomenon existing in Josephson junctions including, for instance, Fraunhofer-like diffraction or a $\pi$-state behavior. 
We also investigate an unconventional model related to inhomogenous Zeeman field caused by magnetic focusing, and note that it can not account for the observed behaviour.
Finally, we consider these results in the context of topological superconductivity, and show that the observed $I_C$ enhancement is compatible with a magnetic field-induced topological transition of the junction.}
\end{abstract}
\maketitle

%INTRO PARAGRAPH ( An introduction (without heading) of up to 500 words of referenced text expands on the background of the work (some overlap with the summary is acceptable), followed by a concise, focused account of the findings, ending with one or two short paragraphs of discussion.)
Coupling a conventional \emph{s}-wave superconductor to materials based on helical electrons such as topological insulators  or semiconductors with strong spin-orbit (SO) interaction like InAs or InSb nanowires (NWs) leads to an unconventional \emph{p}-wave superconductor. 
The latter may undergo a topological transition, and become a topological superconductor (TS) hosting exotic edge states with Majorana-like character~\cite{Beenakker_search_2013,alicea_new_2012}. 
Most of the early experimental efforts to demonstrate these modes have  been focused on normal metal-superconductor junctions realized with strong-SO NWs~\cite{Mourik_Signatures_2012,Das_Zero-Bias_2012,Deng_Parity_2014} 
with the aim to detect signatures of Majorana bound states (MBSs) emerging for increasing Zeeman fields. 
Soon after, experiments on Josephson junctions based on helical materials have  been performed as well to detect peculiar hallmarks of MBSs in the phase evolution of the Josephson effect, both in the context of topological insulators~\cite{Williams_Unconventional_2012,
Hart_Induced_2014,Pribiag_Edge-mode_2015,Sochnikov_Nonsinusoidal_2015,wiedenmann_$4pi$-periodic_2015}
%Qu_Strong_2012,%TI-FR; provide prospects for the realization of devices supporting Majorana fermions
%Veldhorst_Experimental_2012,%TI-SQUID; relevance to the ongoing pursuit of realizing  interferometers for the detection of Majorana fermions in superconductor
%Veldhorst_Josephson_2012,%TI; provide prospects for the realization of devices supporting Majorana fermions
%Oostinga_Josephson_2013, %TI-FR; to search for the manifestation of zero-energy Majorana states in transport experiments.
%Hart_Induced_2014, %TI-FR; as a platform in which to pursue topological superconductivity and Majorana bound states, whether through following existing theoretical proposals or those yet to be formulated
%Kurter_Dynamical_2014, TI-SQUID
%Kurter_Evidence_2015, TI-SQUID
%maier_phase-sensitive_2015,TI-SQUID
%Pribiag_Edge-mode_2015,%TI-2D These experiments establish InAs/GaSb as a promising platform for the confinement of Majoranas into localized states, enabling future investigations of non-Abelian statistics.
%Sochnikov_Nonsinusoidal_2015, %TI-SQUID; These experimental results suggest that the topological properties of the normal state can be inherited by the induced superconducting state, and that 3D HgTe is a promising material for realizing the many exciting proposals that require a topological  superconductor.
%wiedenmann_$4pi$-periodic_2015,%TI-SQUID; Further investigations are required to conclusively demonstrate the relationship of these observations to Majorana physics
and  NWs~\cite{rokhinson_fractional_2012}. 
Yet, a conclusive evidence of MBSs emerging from these hybrid systems still remains  an outstanding experimental goal.
This is particularly true for NW-based Josephson weak links where no signatures of TS have been reported in the supercurrent~\cite{Doh_tunable_2005,De_Franceschi_Hybrid_2010,Nilsson_Supercurrent_2012,Abay_high_2012,Abay_quantized_2013,Paajaste_Pb/InAs_2015,Abay_Charge_2014,roddaro_hot-electron_2010}.

In this work, we investigate the Josephson coupling in Al/InAs-NW/Al hybrid junctions in the presence of an external magnetic field. 
In particular, we focus on the amplitude of the Josephson critical current $I_C$ which is expected to strongly increase when the topology of the junction enables the emergence of MBSs ~\cite{San-Jose_Mapping_2014}.
Our key result is the observation of a colossal \emph{enhancement} of $I_C$ manifesting just above a threshold magnetic field applied in perpendicular direction with respect to the junctions substrate and the NW. 
This is in stark contrast with what has been observed so far in conventional weak links, where the Josephson coupling turns out to be suppressed
by any applied magnetic field.
%In the very first approximation it is well known that a magnetic field destroys superconductivity via the orbital and paramagnetic effect and hence one expects correspondingly a monotonic decay of the critical Josephson current by applying an external field.
Notably, this abrupt switching of $I_C$ always occurs  around the same magnetic field ($B_{sw}$) in junctions made of nominally identical NWs but characterized by different lengths of the normal (N) region, 
therefore suggesting that the origin of the observed critical current enhancement is \emph{intrinsic}, i.e., it is not related to geometrical resonances existing in the junction or linked to the Thouless energy of the system~\cite{furusaki_josephson_1992}. 
In addition, the temperature dependence of $B_{sw}$ follows a BCS-like behavior thus indicating a strong link to the proximity effect present in the junction. 

Our results are discussed on the basis of the most common and understood phenomena that may affect the Josephson  supercurrent of the junction. 
Although no final conclusions can be drawn from such an analysis we stress that a \emph{topological transition} developing at $B_{sw}$ appears to be fully compatible from a qualitative point of view with the observed phenomenology: our proposed physical scenario consists of a magnetically-driven zero-energy parity crossing of Andreev levels in the junction~\cite{San-Jose_Mapping_2014} which is expected to occur for magnetic fields applied perpendicularly to the wire SO axis, exactly as observed in the experiment.

%%%%%%%%%%%%%%%%%% FIGURE 1 %%%%%%%%%%%%%%%%%%

\begin{figure}[t!]
\includegraphics[width=\columnwidth]{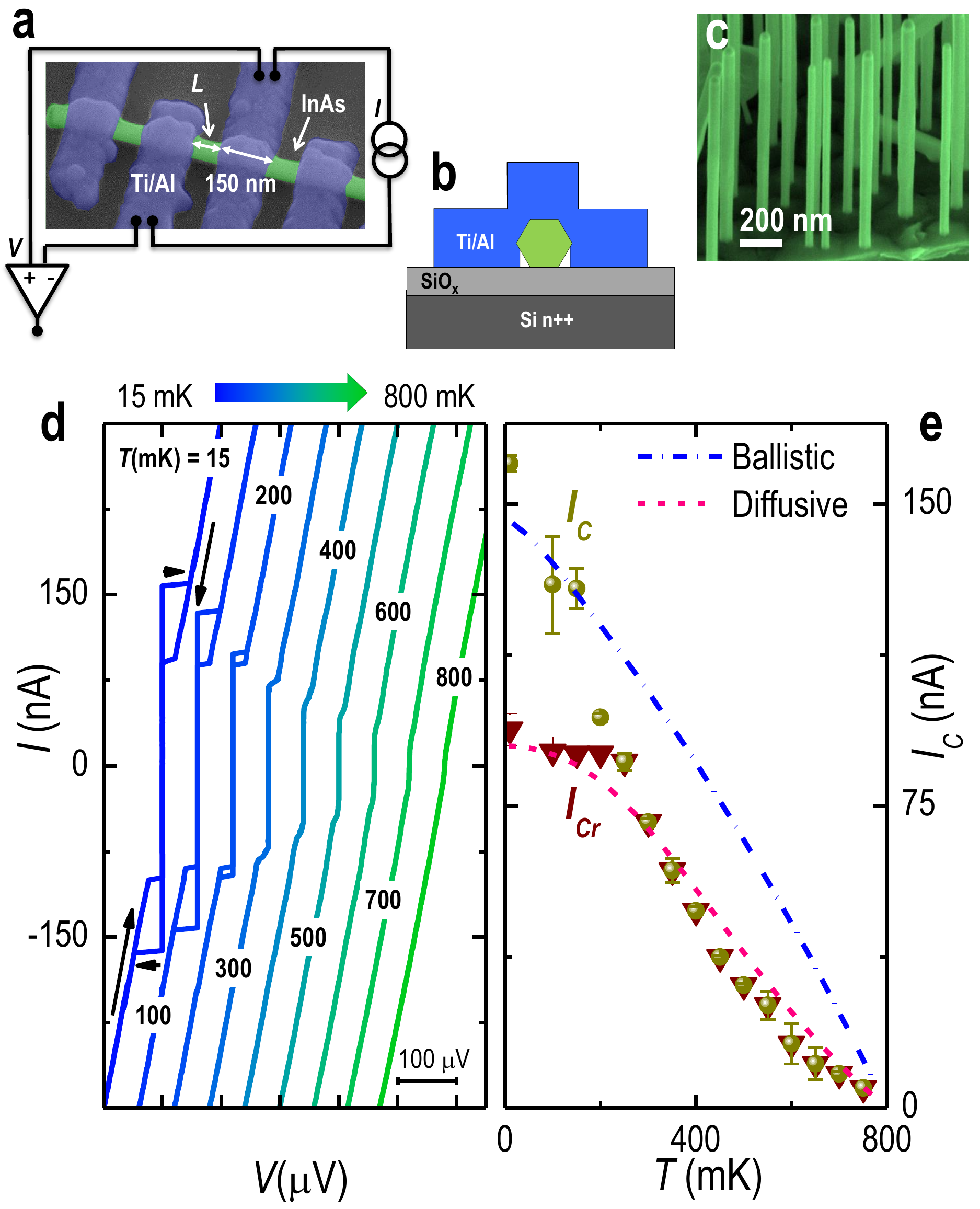}
\caption{\label{NW-device-sketch}
\textbf{\textbar ~Sample layout and zero magnetic field characterization.}
\textbf{a}, Pseudo-color scanning electron micrograph of a typical \emph{n}-InAs nanowire-based Josephson junction along with a sketch of the four-wire measurement setup. The Josephson junction is current biased ($I$) whereas the voltage drop ($V$) is measured through a room-temperature differential preamplifier.
The junction length is denoted with $L$, and  the width of the Ti/Al electrodes is  $\sim150$~nm. 
\textbf{b}, Side view of the junction showing the different materials forming the structure.
\textbf{c}, Colored $45^{\circ}$ tilted scanning electron micrograph of a typical distribution of InAs nanowires after their growth. 
%The image was taken at a 45\textdegree angle. 
\textbf{d}, Back and forth current vs voltage characteristics  of an Ti/Al-InAs Josephson junction with $L=100$~nm measured at different bath temperatures $T$. The curves are horizontally offset for clarity. 
%For $T< 300$~mK quasiparticle heating dominates the transition from superconducting to normal state resulting in a strong hysteretic behaviour which depends on the direction of the bias, as indicated by arrows. 
\textbf{e}, Switching ($I_{C}$, dots) and retrapping  ($I_{Cr}$, triangles) supercurrents vs temperature. Two distinct theoretical models  for the critical current $I_{C}$   holding either in the ballistic (dash-dotted line) or  diffusive (dashed line) regime are shown.}
\end{figure}

%%%%%%%%%%%%%%%
%%%%%%%%%%%%%%%%%% FIGURE 2 %%%%%%%%%%%%%%%%%%
\begin{figure*}[t!]
\begin{center}
\includegraphics[width=0.85\textwidth,clip=]{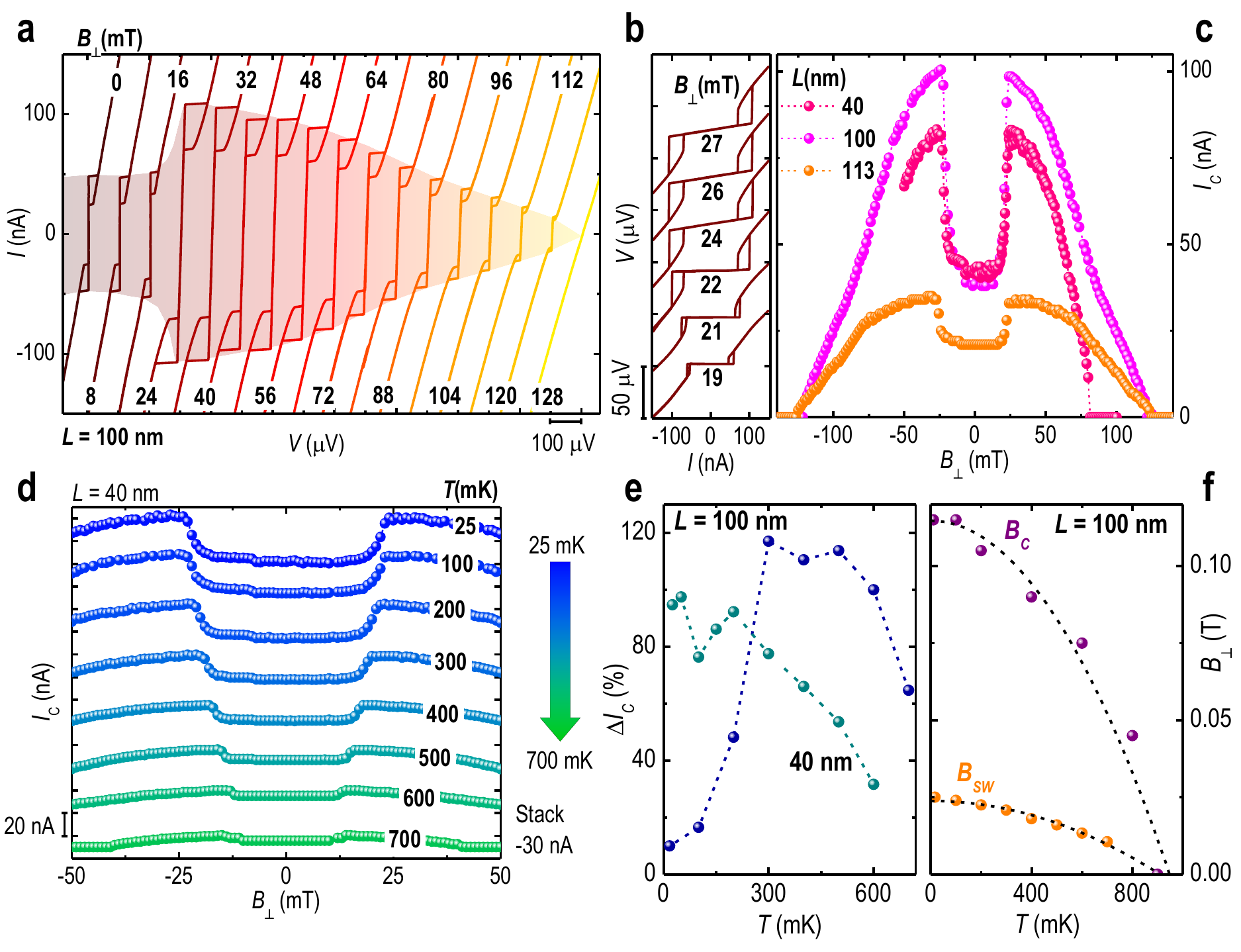}
\caption{
\label{BDependence}
\textbf{\textbar~Enhancement of the critical current at finite out-of-plane magnetic field.}
\textbf{a},~Back and forth $IV$ characteristics of a Josephson junction with $L=100$~nm measured for different values of the out-of plane magnetic field ($B_\bot$) at 15 mK. The curves are horizontally offset for clarity. 
A strong enhancement of $I_C$ occurs at  $B_\bot \simeq 23$~mT.
\textbf{b},~Blow-up of selected $IV$ characteristics of the same junction of panel \textbf{a} showing the dissipative behaviour of the weak link after the transition to the enhanced $I_C$ state. 
\textbf{c},~Comparison of the $I_C$ vs $B_\bot$ behavior for three junctions of different length $L$ at $15$~mK.
\textbf{d},~Critical current $I_{C} $ vs $B_{\bot}$ measured at different bath temperatures for a junction with $L=40$~nm. The curves are vertically offset for clarity. 
Each data point represents the critical current obtained from a single $IV$ measurement at constant magnetic field and temperature.
\textbf{e},~Temperature dependence of the critical current relative enhancement, $\Delta I_C = \text{max}[I_C(B_\bot)]/I_C(0)-1$,
for two different junctions with $L=100$~nm and $L=40$~nm. Note that the junction with $L=100$~nm is different from the one shown in panel \textbf{c}.
\textbf{f},~Temperature dependence of the critical field $B_C$ for the disappearance of the Josephson effect, and of the switching field $B_{sw}$ for  $I_C$ for a junction with $L=100$~nm. Dotted lines are the BCS fitting of the two data sets using the equation $B_x(T)=B_x(0)[1-(T/T_C)^2]$ obtained for the same superconducting critical temperature $T_C=900$~mK.}
\end{center}
\end{figure*}

\subparagraph*{Experimental results}

Figure~\ref{NW-device-sketch}a shows a scanning electron micrograph of a typical \emph{n}-InAs-NW-based Josephson junction. 
The junction's interelectrode  spacing $L$ ranges from $\sim 40$ nm to $\sim 113$ nm.
The growth and physical details of the \emph{n}-doped InAs NWs are reported in the Methods section. 
For each junction a neighbouring Ti/Al superconducting pair is used for transport measurements. 
The current vs voltage ($IV$) characteristics of the Josephson weak links are obtained by applying a bias current $I$, and  measuring the resulting voltage drop across the NW via a room-temperature differential preamplifier, as shown in Fig.~\ref{NW-device-sketch}a. 
A  shematic side view  showing the materials forming the junctions as well as a $45^{\circ}$ tilted scanning electron micrograph of a typical distribution of \emph{n}-InAs NWs after the growth are displayed in  Fig.~\ref{NW-device-sketch}b and c, respectively. 

Figure~\ref{NW-device-sketch}d shows the temperature dependence of the $IV$ characteristics of a typical Josephson junction with  $L=100$~nm. 
A  critical current $I_C$ exceeding $\sim150$~nA is observed at the base temperature of a dilution refrigerator (15~mK)~\cite{Abay_Charge_2014}. 
Without any applied magnetic field $I_C$ persists up to $\sim800$~mK. 
Furthermore, for temperatures below 250~mK a remarkable hysteretic behaviour between the switching ($I_{C}$) and retrapping ($I_{Cr}$) critical currents is observed, as shown in Fig.\ref{NW-device-sketch}d,e. 
This  hysteresis stems from quasiparticle heating within the  NW region  whilst switching from the resistive to the dissipationless regime \cite{Courtois_origin_2008}, and has been often observed in hybrid Josephson junctions independently of the geometry or composition of the weak link~\cite{Gunel_Supercurrent_2012,fornieri_ballistic_2013,Amado_Electrostatic_2013,Paajaste_Pb/InAs_2015,spathis_hybrid_2011,giazotto_josephson_2011}.

The monotonic decay of $I_C(T)$ and the saturation of $I_{Cr}(T)$ at low $T$ is displayed in Fig.~\ref{NW-device-sketch}(e)~\cite{dubos_josephson_2001}
together with the two best fits for $I_C(T)$ obtained by modelling the junction as an ideal diffusive or ballistic NW (red and blue dashed line, respectively). 
The details of the theoretical model for each fit are described in the Methods section. 
None of the two fitting curves can accurately describe the monotonic decay of $I_C(T)$ which is consistent with a junction belonging to an intermediate regime between the two above limits~\cite{Abay_Charge_2014}, i.e., $L\sim l_e$ where $l_e\sim 60$~nm is the electron mean free path estimated  in our InAs-NWs~\cite{Viti_Se-doping_2012,roddaro_hot-electron_2010}. 
Moreover, the diffusive fit suggests an effective junction length of the order of $ \sim 300 $ nm which largely exceeds the interelectrode spacing. 
The actual geometry of the junction supports this observation since the superconducting electrodes cover a considerable portion of the NW. 
In addition, the same fit yieds an estimate for the junction Thouless energy $ E_{th}=\hbar D/L^{2} \approx 160 \mu$eV ($D$ is the diffusion constant in the InAs NW), and for the induced superconducting minigap~\cite{Hammer_density_2007} in the NW, $\Delta \simeq 80 \mu$eV.

As long as the external magnetic field is absent, the behavior of our NW-based Josephson junctions is fully consistent with what has been observed for similar Al/InAs-NW/Al weak links~\cite{Abay_high_2012, Abay_quantized_2013,Abay_Charge_2014,roddaro_hot-electron_2010}. 
A strikingly different and unexpected phenomenology develops when the magnetic field is applied perpendicular to the substrate ($B_\bot$): 
the amplitude of $I_C$ drastically changes in a way that has never been observed so far in similar devices~\cite{Abay_high_2012, Abay_Charge_2014} and, to the best of our knowledge, in any other kind of Josephson weak links.
As shown in Fig.~\ref{BDependence}a, where the $IV$ characteristics of one of the junctions with $L=100$~nm is displayed for different values of $B_\bot$, 
the critical current $I_C$ remains almost constant up to $\sim15\,{\rm mT}$, then quickly doubles its amplitude at a ``switching'' field $B_{sw}\simeq $~23~mT, and decays at larger magnetic fields. 
Furthemore, for $|B_\bot|>B_{sw}$, the $IV$ characteristics develop a dissipative behavior (i.e., they show a finite slope  around $V\approx0$, see Fig.~\ref{BDependence}b) which corresponds to a resistance of $\sim 60 \Omega$.
The colossal enhancement of $I_C$ (which exceeds 100\%) is very robust and reproducible over different cooling cycles, and measured junctions. 

Figure~\ref{BDependence}c shows the behavior of $I_C(B_\bot)$ for three junctions of different lengths. 
Apart from sample specific fluctuations of $I_C(0)$, the supercurrent enhancement occurs at the \emph{same} magnetic field for all the junctions.
This suggests that the origin of the effect is intrinsic to the materials combination, and cannot be attributed to geometrical resonances in the junction~\cite{furusaki_josephson_1992}.
A nontrivial behaviour characterizes also the temperature evolution of the relative $I_C$ enhancement, $\Delta I_C = \text{max}[I_C(B_\bot)]/I_C(0)-1$, shown in Fig.~\ref{BDependence}d,e for two junctions with $L=40$~nm and 100~nm belonging to different NWs. 
In contrast to the usual temperature-driven weakening of the Josephson effect, $\Delta I_C$ shows a nonmonotonic behaviour with a maximum at $T\sim 300$~mK for the junction with $L=$100 nm whereas in the shorter junction ($L=40$~nm) the behaviour is almost monotonic, and simply decays with the temperature. 
Several characterizations performed on different samples seem to indicate that the temperature behaviour of $\Delta I_C$ is not related to the length of the junction but rather depends on the specific NW. 
Moreover,  the behavior of $B_{sw}(T)$ shown in Fig.~\ref{BDependence}f for a junction with $L=100$~nm, follows the same temperature dependence of the critical field for the disappearance of the Josephson effect $B_C$ (also displayed in the same plot) therefore suggesting a common origin of the two phenomena, i.e., the proximity effect.
%%%%%%%%%%%%%%%%%% FIGURE 3 %%%%%%%%%%%%%%%%%%
\begin{figure*}[t!]
\begin{center}
\includegraphics[width=0.8\textwidth,clip=]{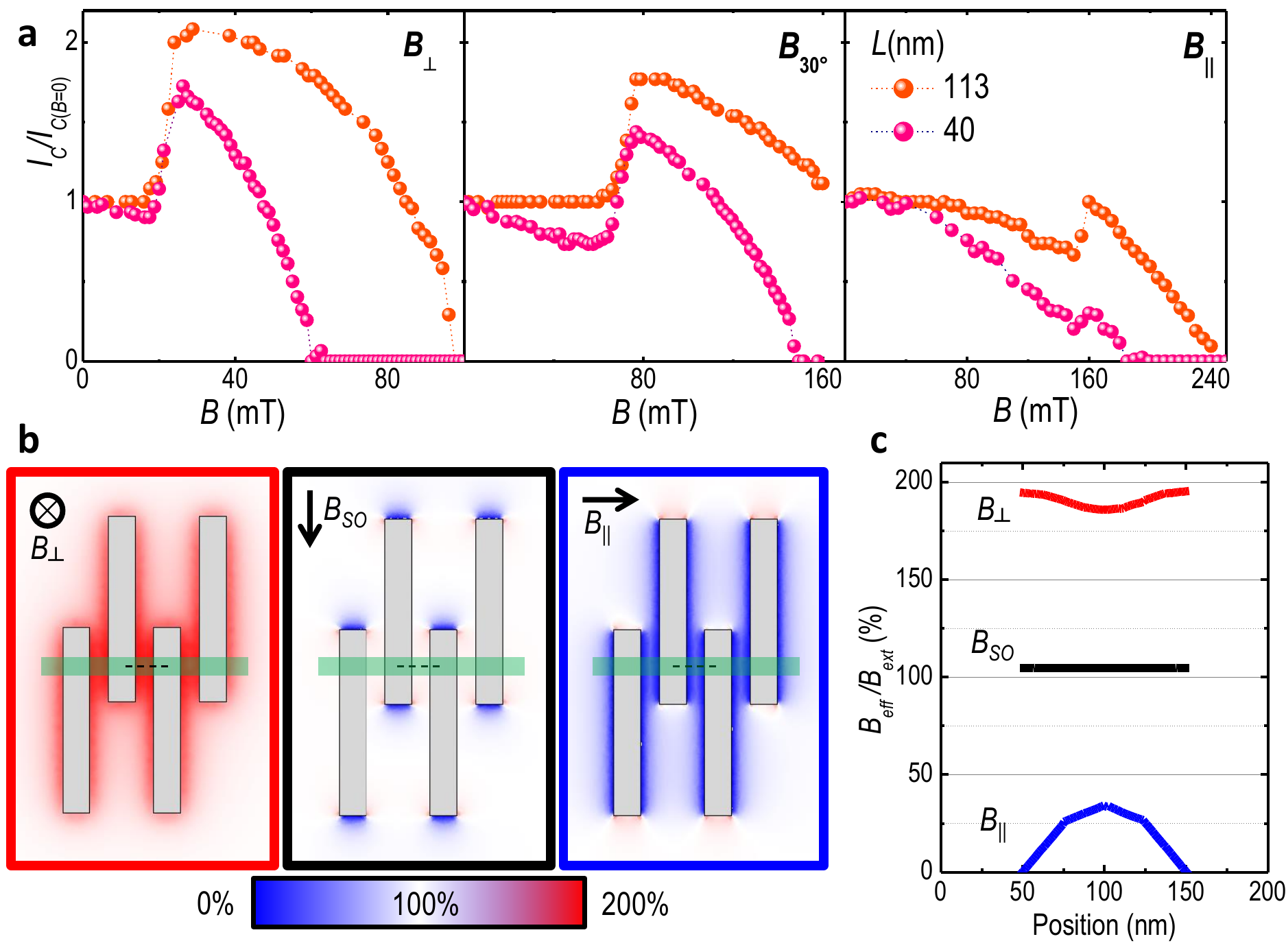}
\caption{
\label{ADependence}
\textbf{\textbar~Angle dependence of the critical current enhancement.}
\textbf{a}, Comparison between the $I_C$ behaviour as a function of the magnetic field applied in three different orientations for two different junction lengths: out-of plane ($B_\bot$), $30^{\circ}$  from the plane ($B_{\text{30\textdegree}}$), and in-plane ($B_{||}$).
\textbf{b}, COMSOL finite-element simulation of the nonuniform distribution of the amplitude of $B$ due to the Meissner effect in the superconducting leads, evaluated for a magnetic field applied along the three main orthogonal axes of the junction: out-of plane ($B_\bot$, left pannel), in-plane orthogonal to the NW and parallel to the SO vector ($B_{SO}$, middle pannel), and along the NW ($B_{||}$, right pannel).
Vertical gray regions indicate the superconducting electrodes whereas the light-green horizontal ones the NW.
\textbf{c}, Amplitude profile of the effective magnetic field ($B_{eff}$) normalized by the applied magnetic field ($B_{ext}$) along the portion of the NW indicated by the dashed lines shown in \textbf{b}. 
Note that while in the out-of-plane direction ($B_\bot$, red line) $B_{eff}$ is almost uniformly doubled  with respect to $B_{ext}$ owing to the focusing effect, along the NW ($B_{||}$, blue line) $B_{eff}$ it is drastically suppressed by Meissner screening,  and reaches at most $\sim25\%$ of the intensity of the external field at the center of the junction. In the in-plane direction orthogonal to the NW ($B_{SO}$, black line) $B_{eff}$ almost coincides with $B_{ext}$.  
}
\end{center}
\end{figure*}

A crucial feature observed in the $I_C$ enhancement of all the junctions, and which is essential in order to discriminate over the possible origins of this effect, is the strong dependence of  $I_C(B)$ on the orientation of the external magnetic field. 
This clearly appears in Fig.~\ref{ADependence}a, where we compare the critical currents of two junctions with different lengths ($L= 40$~nm and 113~nm) measured for three different orientations of $B$. 
The maximum $I_C$ enhancement is observed when the field is applied perpendicular to the substrate (and thus also to the NW axis) as shown in the left panel, and occurs at $B_{sw}\simeq 23$~mT. 
Differently, in canted field configurations ($B_{30^{\circ}}$, central panel) $B_{sw}$ shifts to higher fields according to the amplitude of the projection $B_{\bot}$. 
When $B$ is applied in-plane ($B_{||}$, right panel) the effect is almost absent apart a tiny supercurrent enhancement around $\sim160$~mT which can be ascribed to a small misalignment present in our setup as well as to an incomplete magnetic field screening in the junction region, as will be discussed below.
Notably, the in-plane components of the magnetic field (see right panel) seem to have a marginal role in determining the actual value of $B_{sw}$, this conclusion following from the similar behavior displayed by the two above Josephson weak links which were fabricated with NWs having very different orientations in the substrate plane.

The peculiar dependence of the $I_C$ enhancement as a function of magnetic field direction imposes an important constraint over possible models that can explain the effect. 
These observations joined with the in-plane pinning of the spin-orbit (SO) vector in InAs NWs laying on top of a SiO$_2$/Si substrate~\cite{Mourik_Signatures_2012}, seem to lead to the intriguing conclusion that the $I_C$ enhancement requires the magnetic field to be \emph{perpendicular} to the SO vector. 
The lack of $I_C$ enhancement observed for field configurations orthogonal to the SO vector but parallel to the NW can be attributed to the strong magnetic field expulsion in the weak link region due to the Meissner screening of the superconducting electrodes forming the junction.
The magnitude of this effect has been numerically estimated for our junctions geometry for the three relevant directions of the external magnetic field ($B_{ext}$). 
The results are summarized in Fig.~\ref{ADependence}b where the space distribution of $B$ is calculated assuming an ideal Meissner effect in the superconducting leads. 
In particular, we obtain that when $B_{ext}$ is orthogonal to the substrate ($B_{\bot}$, left panel) the magnetic field $B_{eff}$ in the junction region is strongly amplified, and its intensity is almost doubled  with respect to the external field due to magnetic focusing, as recently reported for Pb-based InAs NW Josephson junctions~\cite{Paajaste_Pb/InAs_2015}. 
When $B_{ext}$ is applied in-plane along the SO vector ($B_{SO}$, central panel) there is complete penetration of the magnetic field within the junction region, i.e., $B_{eff}\sim B_{SO}$.
By contrast, when $B$ is applied parallel to the NW axis ($B_{||}$, right panel) the weak link area is almost screened by the magnetic field thanks to Meissner expulsion in the leads, and $B_{eff}$ obtains values up to $\sim 25\%$ of the external  field at the center of the junction. 
Figure~\ref{ADependence}c summarizes the above results for the amplitude profile of $B_{eff}$ along the portion of the NW indicated by the dashed lines in panel b.  

\subparagraph*{Theoretical interpretations}
The scenario drawn above by these experimental evidences is clear but its interpretation is puzzling due to the complexity of the system and the large amount of non-trivial phenomenologies possible for Josephson junctions.
In the first approximation it is known that a magnetic field destroys superconductivity via the orbital and paramagnetic effect, and therefore one would expect a monotonic decay of the critical  current by the application of an external field.
An experimental exception to this behavior is represented by field-enhanced superconductivity observed in Josephson junctions made with metallic NWs covered with magnetic impurities. 
In that case,  the $I_C(B)$ enhancement is induced by the field polarization of local moments, and by the relative quenching of the exchange coupling with the electrons in Cooper pairs~\cite{Rogachev_Magnetic-Field_2006}. 
Due to the absence of magnetic impurities in our NWs, and owing to the specific magnetic field orientation leading to the $I_C$ enhancement we can exclude this picture as the explanation of our observations.

Two further mechanisms are known that yield an $I_C$ increase as a function of field: Fraunofer-like diffraction~\cite{Cuevas_Magnetic_2007}, and the $\pi$-junction behaviour~\cite{Yokoyama_Anomalous_2014}. 
The former is expected to occur whenever the magnetic flux enclosed by the junction, $\Phi=LWB_\bot$ ($W$ is the NW diameter), equals an integer multiple of the flux quantum, $\Phi_0=h/2e$. 
Note that neither the absence of an $I_C$ maximum around $B_\bot=0$ nor the evidence that the critical current enhancement is independent of the junction length can be explained within this picture. 
On the other hand, a $\pi$-junction behavior induced by the external field is typically characterized by oscillations of $I_C(B)$ with periodicity of the order $\sim g\mu_B B/E_T$, i.e., the ratio between the Zeeman and the Thouless energy of the junction, where $g$ is the NW gyromagnetic factor and $\mu_B$  is  the Bohr magneton.
Such oscillatory behaviour, when combined with SO coupling and disorder~\cite{Yokoyama_Anomalous_2014}, might explain enhancement of the critical current for some specific values of the magnetic field. 
However, neither the estimated value for the ratio $g\mu_B B_{sw}/E_T\sim 0.25$ nor the absence of any dependence on the junction length (through the Thouless energy $E_{T}$)
%(through the relation existing between the Thouless energy and the length in a ballistic junction $E_T=\frac{\hbar v_F}{L}$) 
can be accommodated within this explanation. 
%%%%%%%%%%%%%%%% FIGURE 4 %%%%%%%%%%%%%%%%%%
\begin{figure}[t!]
\centerline{\includegraphics[width=\columnwidth,clip=]{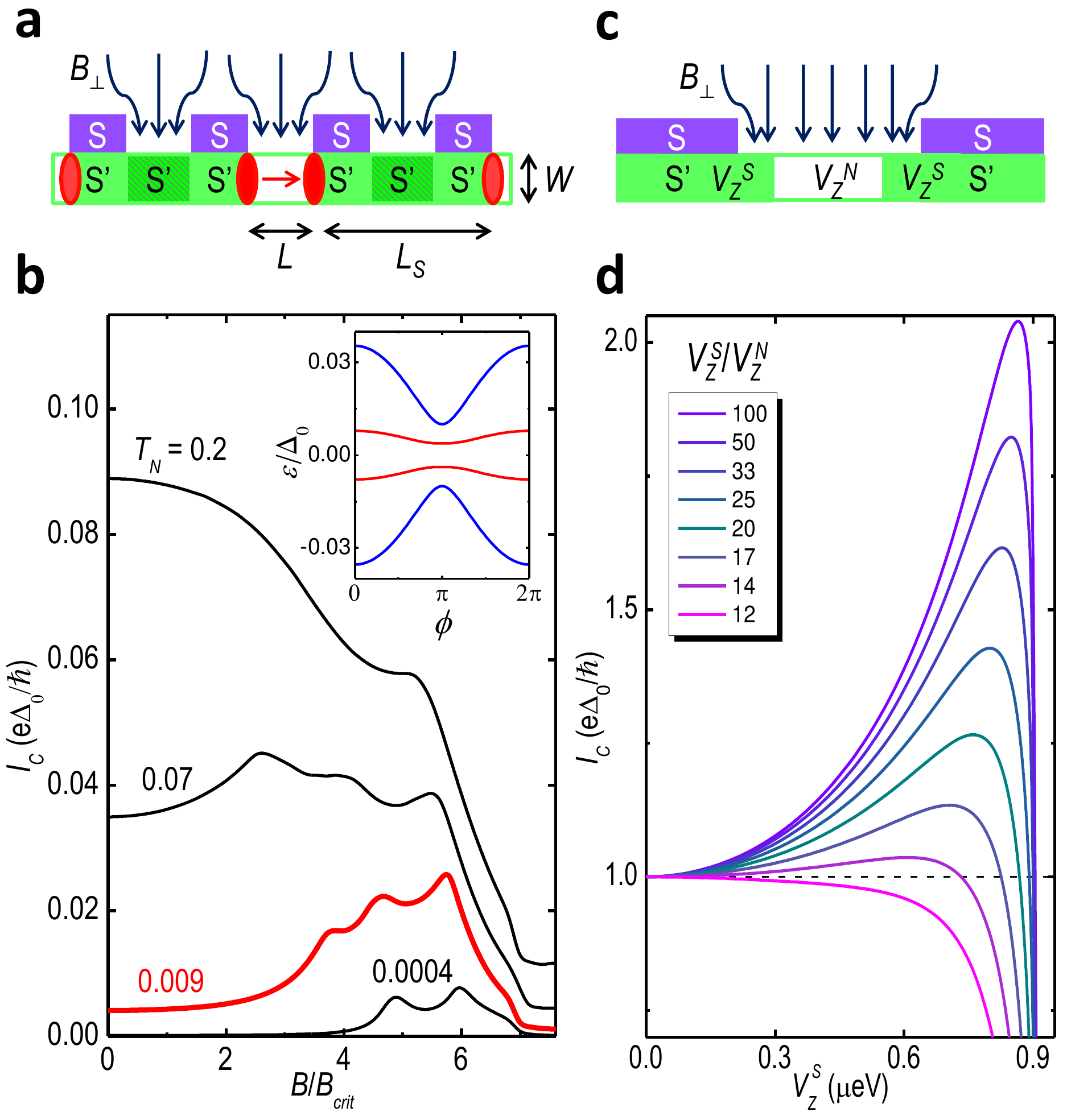}}
\caption{\label{Model1}
\textbf{\textbar~Two possible models for the $I_C$ enhancement.}
\textbf{a},~A sketch of the four Majorana bound states (red circles) formed in the two proximized regions S' of a (\emph{ballistic}) NW under the action of an out-of-plane magnetic field $B_\bot$.
\textbf{b},~Theoretical critical current $I_C$ calculated for increasing magnetic  field values in an InAs  Josephson junction similar to the ones investigated in the experiment ($L=50$~nm, $L_S=500$~nm, and $W=50$~nm). 
The NW chemical potential $\mu$ is set to have six spinful subbands at half filling. 
Rest of the parameters: spin-orbit coupling $\alpha_\mathrm{SO}= 0.13 \,\mathrm{eV\AA}$, gyromagnetic factor  $g=15$, and zero-temperature superconducting energy gap $\Delta=150\,\mu$eV. 
The red curve is qualitatively representative of the $I_C$ enhancement observed in the experiment. 
$T_N$ denotes the contact transmissivity.
Inset: corresponding Andreev levels $\varepsilon$ vs phase difference $\phi$ at $B = 6.2 B_{crit}$, which exhibit an avoided crossing at phase difference $\phi=\pi$ due to the hybridization of inner (blue) and outer (red) Majorana bound states across $L_S$.
\textbf{c},~Sketch of a (\emph{diffusive}) NW Josephson junction in the presence of an inhomogeneous magnetic field due to the focusing effect induced by the two superconducting leads.
The local Zeeman exchange energy in the superconductors S' and in the NW is denoted by $V^{S}_Z$ and $V^{N}_Z$, respectively, with $V^{S}_Z>V^{N}_Z$. 
\textbf{d}, Theoretical critical current $I_C$ calculated as a function of $V^{S}_Z$ for different degrees of inhomogeneity ($V^{S}_Z / V^{N}_Z$).
}
\end{figure}

Despite being a rather rare phenomenon in Josephson junctions we then consider two less obvious alternatives which account, at least from a qualitative point of view, for the observed colossal critical current enhancement induced by a magnetic field. 
To embrace both transport regimes which are relevant for our junctions, we will present: (i) a first model based on topological transitions in the ballistic approximation; (ii) a second model based on inhomogeneous Zeeman field in the diffusive limit.

\paragraph*{(i) Ballistic scenario -Topological transition and MBSs-} The enhancement of the critical current in a NW-based superconductor-normal metal-superconductor Josephson junction was predicted to occur after the proximized sections of the nanowire (the S' regions shown in  the scheme of Fig.~\ref{Model1}a) are driven into a topologically non-trivial phase by an external Zeeman field. 
In particular, it was shown~\cite{San-Jose_Mapping_2014} that in the topologically non-trivial phases the critical supercurrent of a Josephson junction realized with a multimode quasi-one dimensional semiconductor NW with SO (Rashba-type) coupling, like the ones studied here,  can be strongly enhanced relative to the trivial phase for small junction transmissivity ($T_N$). 
This happens by virtue of the additional supercurrent contributed by Majorana zero modes in the junction as the external Zeeman field exceeds a critical value ($B_{crit}$). 
In a quasi-one dimensional geometry, this topological transition occurs as a given subband $n$ undergoes a band inversion at Zeeman energy $V^{(n)}_Z=\sqrt{\Delta^2+\mu_n^2}$, where $\mu_n$ is the Fermi energy measured from the bottom of the subband, and $\Delta$ is the superconducting minigap induced in the NW. 
After the transition, and for long enough proximized regions, i.e., for $L_S\gg\xi\equiv \hbar v_F/\Delta$ where $\xi$ is the Majorana state localization length, each S' section of the NW becomes a topological superconductor with emergent Majorana zero modes at its ends (the red circles in Fig.~\ref{Model1}a).  
The topological transition can be directly seen as a closing and reopening of the Boboliubov-De Gennes (BdG) spectrum near zero energy. 
After the transition, the BdG spectrum contains emergent Majorana zero modes with protected crossings at a superconducting phase difference $\phi=\pi$ which give rise to a $4\pi$-periodic Josephson effect, and to an enhanced critical current ($I_C\sim\sqrt{T_N}$) relative to that of conventional Andreev levels ($I_C\sim T_N$) therefore only observable for reduced contact transmissivity $T_N\ll1$. 
For shorter wires (with $L_S\ll\xi$) the Majorana modes are always strongly overlapping, and merge into standard finite-energy Andreev levels. 

Figure.~\ref{Model1}b illustrates this phenomenon displaying the calculated spectrum of a NW Josephson junction with $L_S=500$~nm corresponding to the experimental samples (note that we take into account all the S fingers on each side of the junction). 
In this limit the junction should be always topologically trivial since the protected $\phi=\pi$ crossing is lifted, as shown in the inset of Fig.~\ref{Model1}b. 
Despite this, the $I_C$ enhancement predicted to occur at the topological transition for ideal junction lengths ($L_S\gg\xi$), remains visible even in our short-contact geometry ($L_S\ll\xi$) as a reminiscence of the Majorana zero modes which, even if strongly overlapping, are effectively contributing to $I_C$. 
This is shown in Fig.~\ref{Model1}b where a remnant of a topological transition for short $L_S$ leads to a critical current enhancement around a crossover magnetic field $V^{(n)}_Z\gtrsim\Delta$ (for $\mu_n\sim 0$) as the junction transmissivity is reduced. 
For the typical energy gap deduced from the experiment, $\Delta\sim 80\,\mu$eV, and assuming $g=15$ for InAs NWs, this yields a critical magnetic field $B_{crit}=\Delta/(\frac{1}{2}g\mu_B)\simeq 184$~mT. 
This latter value is in rough agreement with the experimental magnetic fields at which the $I_C$ enhancement is observed, $B\approx 23$~mT, if we consider the field focusing factor of 2 (as shown in Fig.~\ref{ADependence}c) and a large $g$ factor, $g\sim 60$.

\paragraph*{(ii) Diffusive scenario -Inhomogeneous Zeeman field-}
If we exclude the above topological transition scenario as the origin of the observed effect an alternative model to describe the colossal $I_C$ enhancement is
based on the increase of the critical current predicted to occur in  SF-I-SF Josephson junctions, where SF denotes a superconductor with a Zeeman-split density of states, and I  is an insulator. 
If the effective spin-splitting Zeeman fields in the left and right electrodes are different, $V^{l(r)}_Z$, one expects to observe an enhancement of the critical current with a maximum at $V^{l}_Z-V^{r}_Z=\Delta_{l}+\Delta_{r}$~\cite{bergeret_enhancement_2001,chtchelkatchev_josephson_2002} as a consequence of a robust spin triplet component in the superconducting condensate, where $\Delta_{l(r)}$ is the pairing potential in the left(right) superconductor. 
It is worthwhile to emphasize that the $I_C$ enhancement only takes place if the transmissivity of the barrier is low \cite{chtchelkatchev_josephson_2002} 
to avoid  the coupling between  the two spin-polarized superconducting condensates.
In addition, the critical current enhancement is also possible in SF-I-N-I-SF \cite{Strambini_Mesoscopic_2015} and even in S-I-N-I-S junctions, the latter being a geometry  somewhat closer to the experimental one, if the following conditions are fulfilled: 
(i) The density of the states of the S electrodes is BCS like with the coherent peaks Zeeman split thanks to the presence of the magnetic field (the orbital effects are disregarded); 
(ii) The Zeeman fields in the S and N region ($V^{S(N)}_Z$) have to be very different in amplitude; specifically, it is required $V^{S}_Z\gg V^{N}_Z$
to achieve the observed $I_C$ enhancement. 
(iii) The interface between the S electrode and the N wire should exhibit a tunnel-like transmissivity.
Condition (i) is plausible for the regions of the NW close to the S/N contacts, therefore one can identify the structure as a SF-I-N-I-SF, where now SF denotes a BCS superconductor with a variable spin-splitting proportional to the applied magnetic field.  
Condition (ii), namely an inhomogeneously distributed Zeeman field, is  possible as well owing to the strong magnetic field focusing originating from the S electrodes (see Figs.~\ref{ADependence}b,c) so one can safely assume that $V^{S}_Z\neq V^{N}_ZS$ (see the scheme in Fig.~\ref{Model1}c). 
Condition (iii), on the other hand, is more unlikely but still possible by considering local defects in the NW. 

Having in mind these limitations, we have computed the critical current of a diffusive SF-I-N-I-SF junction with the help of the quasiclassical Green's function formalism~\cite{Strambini_Mesoscopic_2015} (see Methods for more details), and using the parameters obtained from the $I_C(T)$ fit (see Fig.~\ref{NW-device-sketch}e). 
The results are summarized in Fig.~\ref{Model1}d. 
The overall behaviour is qualitatively consistent with the experimental observation although a quantitative comparison is difficult due to the complexity of the junction geometry. 
Despite the good qualitative description of $I_C(B)$, the main discrepancy between the model and the experiment lies in the large value of field inhomogeneity, $V^{S}_Z / V^{N}_Z\gtrsim 10$, required to obtain a  magnetically-driven $I_C$ enhancement. 
This last condition is indeed much stronger than the maximal expectation obtained for our junctions geometry in the out-of-plane magnetic field configuration, i.e., $V^{S}_Z / V^{N}_Z \lesssim 1.2$, as displayed in Fig.~\ref{ADependence}c. 

\subparagraph*{Conclusions}
We have reported a colossal enhancement of the critical supercurrent at finite magnetic fields in \emph{n}-InAs NW-based weak links which is 
in total contrast with the behavior observed so far in conventional Josephson junctions. 
The effect manifests itself  only for very specific experimental conditions: 
the magnetic field needs to be applied perpendicular to the NW axis and to the substrate, i.e., in the \emph{only} configuration which is expected to be perpendicular to the SO vector in the weak link. 
Despite this clear but puzzling experimental evidence, a conclusive theory capable to fully describe our results is still missing. 
We presented two possible models that allow to qualitatively grasp the observed effect, one holding in the ballistic limit, the other in the diffusive regime. 
The available evidences fall in a complex intermediate regime that prevents a detailed agreement between theory and experiment. 
Yet, the realization of fully ballistic junctions is expected to provide an improved  understanding of the present phenomenology. 
In addition, extension of the theory of topologically non-trivial pairing to the case of diffusive NW Josephson weak links will offer a refined description of our system, and further clarify the role of MBSs in the experiment. 
We believe that these newsworthy results will stimulate the development of novel theoretical models, and will contribute to the progress of the investigation and understanding of MBSs  in condensed matter physics.

\section*{ACKNOWLEDGEMENTS}
Partial financial support from the Marie Curie Initial Training
Action (ITN) Q-NET 264034, from the MIUR-FIRB2013 Project Coca
(Grant No. RBFR1379UX), from CNR through the bilateral CNR-RFBR project 2015-2017,
and from the European Research Council
under the European Union's Seventh Framework Programme
(FP7/2007-2013)/ERC Grant 615187-COMANCHE is acknowledged.
The work of E.S. is funded by the Marie Curie Individual Fellowship MSCA-IFEF-ST No. 660532-SupeMag. 
M.A. acknowledges the support of the Tuscany region through the project 'TERASQUID' and the People Programme (Marie Curie Actions) of the European Union H2020 Programme under REA grant agreement n° [656485]. 
The work by P.S-J. is supported by the Spanish Ministry of Economy and Innovation through Grant No. FIS2011-23713 and the Ram\'on y Cajal programme.
The work of R.A. is funded by the Spanish Ministerio de Econom\'{i}a y Competitividad through grant FIS2012-33521.
The work of F.S.B is supported by the Spanish Ministerio de Econom\'{i}a y Competitividad through the Project No. FIS2014-55987-P and Grupos Consolidados UPV/EHU del Gobierno Vasco (Grant No. IT-756-13).

\section*{Methods}
\subparagraph*{Device fabrication}
The Josephson junctions presented in this work are based on heavily \textit{n}-doped InAs NWs grown by metal-assisted chemical beam epitaxy~\cite{Ercolani_InAs/InSb_2009}. NWs are grown in a Riber C-21 reactor by using metallic seeds obtained from thermal dewetting of a thin Au film layer evaporated on a InAs substrate \cite{gomes_controlling_2015,Paajaste_Pb/InAs_2015}. 
Trimethylindium (TMIn) and tertiarybutylarsine (TBAs) (cracked at 1000\textdegree~C) are used in the growth as metallorganic precursors while ditertiarybutyl selenide (DtBSe) is used as a selenium source for \textit{n}-type doping. 
Based on previous experiments performed on similar NWs \cite{Viti_Se-doping_2012,Paajaste_Pb/InAs_2015} we estimate a carrier density $n_{s} \sim 3 \times 10^{18} $ cm$^{-3}$ and an electron mobility $\mu \sim 2000$ cm$^{2}$/Vs from which we deduce a Fermi velocity $v_F \sim 2.2 \times 10^{6}$m/s, an electron mean free path $l_e \sim 60$nm and the effective electron mass for InAs NWs $ m^{*}=0.023m_{e}$, where $ m_{e} $ is the free-electron mass.

After the growth, the NWs are transferred mechanically onto a SiO$_2$/\textit{n}-Si commercial substrate pre-patterned with Ti/Au pads and alignment markers which are defined by optical and electron beam lithography, and deposited by thermal evaporation. 
The position of the NWs on the substrate is mapped with a scanning electron microscope, and used for the aligned electron beam lithography of the Josephson junctions. 
Prior to the deposition of the Ti/Al superconducting electrodes the NWs are etched with a highly-diluted ammonium polysulfide (NH$_{4}$)$_{2}$S$_{x}$ solution to remove the native oxide layer present on the semiconductor surface. 
This procedure improves the quality of the ohmic contact, limiting undesired surface scattering processes. 
The deposition of the Ti/Al (12/78 nm) leads is performed at room temperature in ultra-high vacuum conditions by electron beam evaporation ~\cite{Paajaste_Pb/InAs_2015}.
More than 10 junctions were fabricated starting from five different NWs, and measured at low temperature.

The magneto-electric characterization of the InAs-NW Josephson junctions was performed in a filtered dilution refrigerator down to 15~mK using
a standard 4-wire technique. 
%Even if four superconducting electrodes are present in the devices, only neighbouring pairs are used during the transport measurements. 
The current-voltage characteristics of the junctions were obtained by applying a low-noise biasing current, with voltage across the NW
being measured by a room-temperature battery-powered differential preamplifier.

\subparagraph*{Fitting details of $I_{C}(T)$.}
To study the decay of $I_{C}(T)$ presented in Fig.~\ref{NW-device-sketch}e and identify the main transport regime holding in the NW we have modelled the Josephson junction in two opposite limits: diffusive and ballistic. In the \emph{diffusive} regime, $I_{C}(T)$ is fitted with the expression of the critical current of a superconductor-normal metal-superconductor (SNS) junction obtained  by solving the linearized Usadel equation \cite{Kuprianov_Influence_1988} 
\begin{equation}
\label{diffusive}
I_C=\frac{\pi k_{B} T}{eR_{NW}} \sum_{\omega}\frac{(\kappa_\omega L)\Delta^2  }{ (\omega^2+\Delta^2)\left[\alpha\sinh(\kappa_\omega L)+\beta \cosh(\kappa_\omega L)\right]},
\end{equation}
where the sum is over the Matsubara frequencies
$\omega=\pi k_{B}T(2n+1)$, $n=0,\pm1,\pm2, ...$, $ k_{B} $ is the Boltzmann constant, $\kappa_\omega=\sqrt{2|\omega|/(\hbar D)}$, \textit{D} is the NW diffusion coefficient, $ \hbar $ is the reduced Planck constant, $R_{NW}$ is the resistance of the NW of length \textit{L}, $ \alpha=1+r^2(\kappa_\omega L)^2$, $ \beta=2r(\kappa_\omega L) $, and $r=R_b/R_{NW}$ with $R_b$ being the resistance of the SN interface. 
The best fit with the diffusive model presented in  Fig.~\ref{NW-device-sketch}e (pink dashed line) is obtained  from Eq.~\ref{diffusive} by using $L=300$~nm, $D=0.0416$~m$^{2}$/s, $R_b=4 \Omega $, and $R_{NW}=741 \Omega$.
%The diffusive fit indicates that the effective length of the junction ($ \sim 300 $ nm) is much larger than the interelectrode spacing. The geometry of the device supports this observation, since the electrodes cover a considerable section of the NW. %Moreover from the fit we can estimate the junction Thouless energy to be $ E_{th}=\hbar D/L^{2} \approx 290 \mu$eV and the amplitude of the superconducting minigap induced in the NW that in the diffusive approximation~\cite{Hammer_density_2007} is $\Delta^* \simeq 80 \mu$eV.

On the other hand, for the fit in the \emph{ballistic} regime we use the expression of  the Josephson current $I_J$ valid for a multichannel junction~\cite{beenakker_Universal_1991}
\begin{equation}
I_J(\phi)= \frac{e\Delta(T)^2 }{2\hbar} \sin{\phi} \sum_{n=1}^{N}\frac{D_n}{E_n}\tanh\left[{\frac{E_n(T)}{2k_B T}}\right],
\end{equation}
where $D_n$ are the eigenvalues of the transmission matrix describing the junction, $N$ is the number of conducting channels, $E_n(T) = \Delta(T) \sqrt{1-D_n \sin^2{\phi/2}}$, $\Delta(T)$ is the temperature-dependent BCS energy gap, and $\phi$ is the macroscopic quantum phase difference over the junction. 
The critical current at each temperature is then obtained by maximizing $I_J(\phi)$ with respect to $\phi$, $I_C= \max|_{\phi}[I_J(\phi)]$.
The best fit with the ballistic model shown in  Fig.~\ref{NW-device-sketch}e (blue dashed  line) is obtained by setting $\Delta_0= 120 \mu$eV, $D_n = 1$, and $N = 5$.  

\subparagraph*{Model for the $I_C(B)$ enhancement by a topological transition.}

The Bogoliubov-de Gennes Hamiltonian of a proximized two-dimensional Rashba semiconductor reads
\begin{eqnarray}
H&=&\left(\frac{p^2}{2m^*}-\mu\right)\tau_z+\frac{\alpha_\mathrm{SO}}{\hbar}\left(\sigma_y p_x\tau_z-\sigma_x p_y\right)\nonumber\\
&+&\Delta\sigma_y\tau_y+V_Z\sigma_x\tau_z,
\label{H}
\end{eqnarray}
where $p^2=p_x^2+p_y^2$, $m^*$ is the effective mass, $\sigma_i$ are the spin Pauli matrices, $\tau_i$ are Pauli matrices in the electron-hole sector, and $\alpha_\mathrm{SO}$ is the Rashba spin-orbit coupling. 
The last two terms describe an induced superconducting pairing of strength $\Delta$ and the Zeeman energy $V_Z$ produced by an external magnetic field. It has been shown 
(for details see, e. g.,~\cite{Beenakker_search_2013,alicea_new_2012}) that the \emph{s}-wave pairing in Eq. (\ref{H}) induces both an effective  $p_x\pm ip_y$ intraband pairing, and an interband \emph{s}-wave pairing when reexpressed in terms of the $\pm$ eigenbasis of the helical Rashba + Zeeman normal problem. 
In terms of these effective pairings, the system is topologically non-trivial when only one of the two $p$-wave pairings develops. This occurs when the Zeeman energy $V_Z$ exceeds a critical value. In a quasi-1D geometry (with discrete confinement subbands), this critical value reads
\begin{equation}
V^{(n)}_Z=\sqrt{\Delta^2+\mu_n^2},
\end{equation}
with $\mu_n$ the Fermi energy measured from the bottom of the subband. If the subbands are coupled, through the transverse SO coupling, an even number of Majorana zero modes per edge hybridise in pairs  and form $\lfloor N /2 \rfloor$ full fermions (standard Andreev levels) at non-zero energies $\epsilon_0^{(n)}$ (with $n=1,\dots \lfloor N /2 \rfloor$) localised at each edge. The remaining  $N~\mathrm{mod}~2$ end states stay as Majorana zero modes, at $\epsilon_0^{(0)}=0$. 

This spectral even-odd effect is therefore a manifestation of changes in the topological order in the multiband system, and is reflected in the Josephson effect. In particular, the additional supercurrent contributed by Majorana zero modes in the junction in the non-trivial regime directly maps the topological phase diagram of proximized multiband quasi-one dimensional semiconducting nanowires, as demonstrated in Ref.~\cite{San-Jose_Mapping_2014}.

\subparagraph*{Model for the $I_C(B)$ enhancement by an inhomogeneous Zeeman field.}
We have calculated the critical current $I_C$ of a diffusive SF-I-F'-I-SF 
junction by solving the Usadel equation within the quasiclassical Green's functions formalism~\cite{Strambini_Mesoscopic_2015}.
Above, SF denotes a superconductor (S') Zeeman split by an exchange field $h_{1}$, F' is a N wire possessing an internal exchange energy $h_{2}$, and I represent an insulating tunnel barrier (see the sketch shown in Fig.~\ref{Model1}c). 
$I_C$ can be expressed as
\begin{equation} 
I_C=\frac{\pi A k_{B} T}{e}\sum_{\omega}\left[ \frac{(f^{+})^{2}}{\text{sinh}(\chi^{+}L)\chi^{+}}+\frac{(f^{-})^{2}}{\text{sinh}(\chi^{-}L)\chi^{-}}\right], 
\end{equation}
where $\chi^{\pm}=\sqrt{2(\vert\omega\vert\pm i\text{sign}[\omega h_{2}])/D}$, 
$f^{\pm}=\Delta(h_{1})/\sqrt{(\omega \pm ih_{1})^{2}+\Delta^{2}(h_{1})}$, the sum is over the Matsubara frequencies $\omega$ already introduced, and $\Delta(h_1)$ is the effective superconducting order parameter calculated self-consistently from the BCS gap equation~\cite{giazotto_superconductors_2008}.
Moreover, $A=\frac{ R_{NW} A_{NW}}{ 2R_{C}^2 A_{C}^2 L}$, where $R_{NW}$ is the resistance of the NW of length $L$, $A_{NW}$ is the NW crossection, $R_{C}$ is the contact resistance, and $A_{C}$ is the contact area. For the calculation of the curves displayed in Fig.~\ref{Model1}d we have used the junction geometrical dimensions, the parameters obtained from the diffusive $I_C(T)$ fit shown in Fig.~\ref{NW-device-sketch}e, and $T=60$ mK.
%\bibliographystyle{naturemag_NoURL}
%\bibliographystyle{apsrev4-1}
%\bibliography{Libreriapersonale}

\end{document}